\documentclass{edp-conf}
\usepackage{graphicx}
\def\eg{{\it e.g.} }
\def\etal{{\em et al.} }
\def\ie{{\em i.e.} }

\def\cm2{cm$^2$ }

\def\arcsec{\hbox{$^{\prime\prime}$} }

\begin{document}

\TitreGlobal{SF2A 2002}

\title{SIMBOL--X, an X--ray telescope \\ for the 
0.5--70~keV range} 
\author{Philippe Ferrando}\address{DSM/DAPNIA/Service d'Astrophysique, CEA/Saclay, F91191 
Gif-sur-Yvette France.}

\runningtitle{SIMBOL--X}
\setcounter{page}{237}
\index{Ferrando, P.}

\maketitle
\begin{abstract} 
SIMBOL--X is a high energy ``mini" satellite class mission 
that is proposed by a European collaboration for a launch 
in 2009. SIMBOL--X is making use of a classical X--ray mirror, of 
$\sim$~600~\cm2 maximum effective area, with a 30~m focal length in 
order to cover energies up to several tens of keV. This focal length 
will be achieved through the use of two spacecrafts in a formation 
flying configuration. This will give to SIMBOL--X unprecedented 
spatial resolution (20\arcsec HEW) and sensitivity in the hard X--ray 
range. By its coverage, from 0.5 to 70~keV, and sensitivity, SIMBOL--X 
will be an excellent instrument for the study of high energy processes 
in a large number of sources, both compact and extended.
\end{abstract}
%
\section{Scientific motivation}
The study of the non thermal component in high energy astrophysics 
sources is presently hampered by the large gap in spatial resolution 
and sensitivity between the X--ray and $\gamma$--ray domains. Below 
$\sim$~10~keV, astrophysics missions like XMM--Newton and Chandra are 
using focusing optics based on grazing incidence mirrors giving
extremely good spatial resolution, down to 0.5\arcsec for Chandra, 
and sensitivity. This technique has however an energy limitation at 
$\sim$~10 keV due to the maximum focal length that can fit in a single 
spacecraft. Hard X--ray and $\gamma$--ray imaging instruments, such as 
those on INTEGRAL, are relying on non focusing techniques which do not 
allow to reach spatial resolutions better than $\sim$~10 arc minutes 
and yield much lower point source sensitivities than in low energy 
X--rays.

This transition of techniques unfortunately happens roughly at the 
energy above which the identification of a non thermal component is 
unambiguous with respect to thermal emission. Considered from the low 
energy side, this strongly limits the interpretation of the high 
quality X--ray measurements, and particularly that related to the 
acceleration of particles. Considered from the high energy side, 
this often renders impossible the identification of the source of the 
$\gamma$--ray emission.

The SIMBOL--X motivation is to elucidate the origin of the non thermal 
emission in accretion / acceleration astrophysical sites, both 
compact and extended, by offering a spatial resolution and a 
sensitivity of the ``soft X--ray type" from 0.5 to up to $\sim$ 
70~keV. With SIMBOL--X it will be possible to solve issues or make 
new advances on the physics of the accretion onto black holes (both 
of stellar mass with new capabilities regarding low luminosity states, 
and much more massive as SgrA* or those in AGNs), on the high energy emission 
in jets both of quasars and micro-quasars, on the origin of the hot 
component at the centre of our Galaxy and in clusters of galaxies, on 
the particle acceleration in supernovae remnants, on magnetically 
active stars, or on the diffuse X--ray background in the energy range
where its spectrum is peaking.

\section{Mission concept}
\begin{figure}[hb]
\begin{minipage}{0.49\linewidth}
\centering
\includegraphics[width=\linewidth]{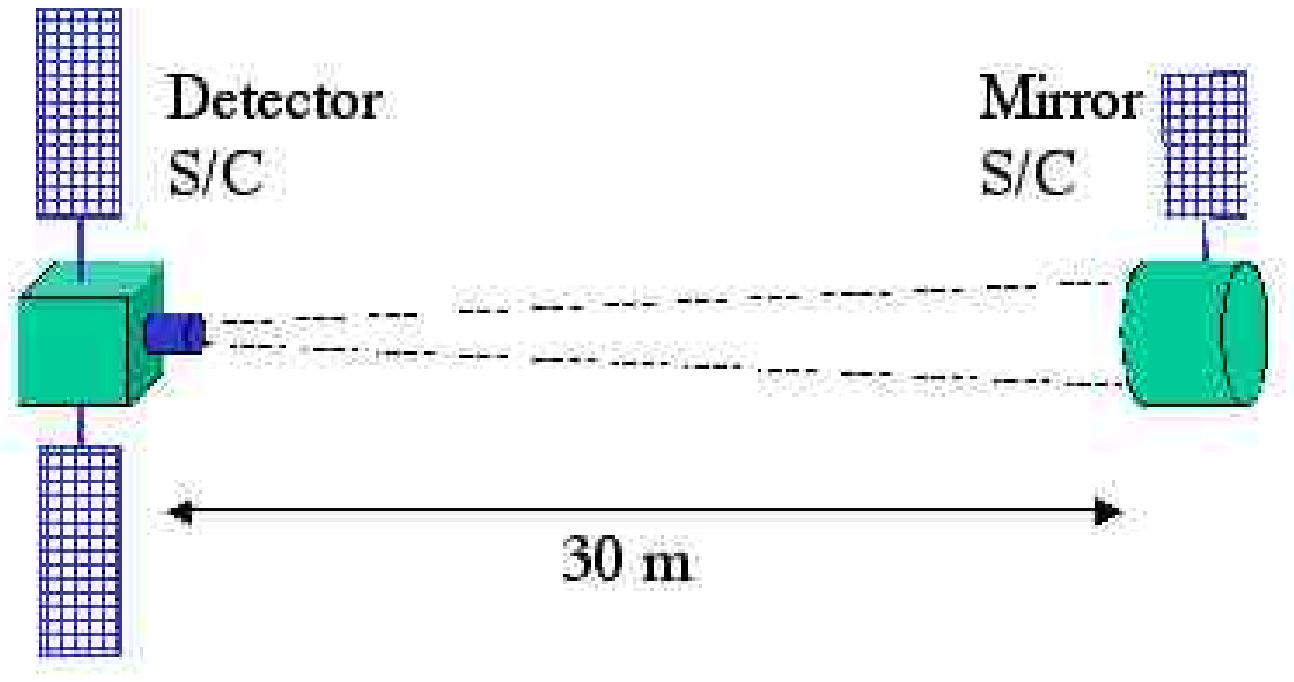}
\caption{SIMBOL--X two spacecrafts configuration.}
\label{fig:fly}
\end{minipage}
\hfill
\begin{minipage}{0.49\linewidth}
\centering
\includegraphics[width=\linewidth]{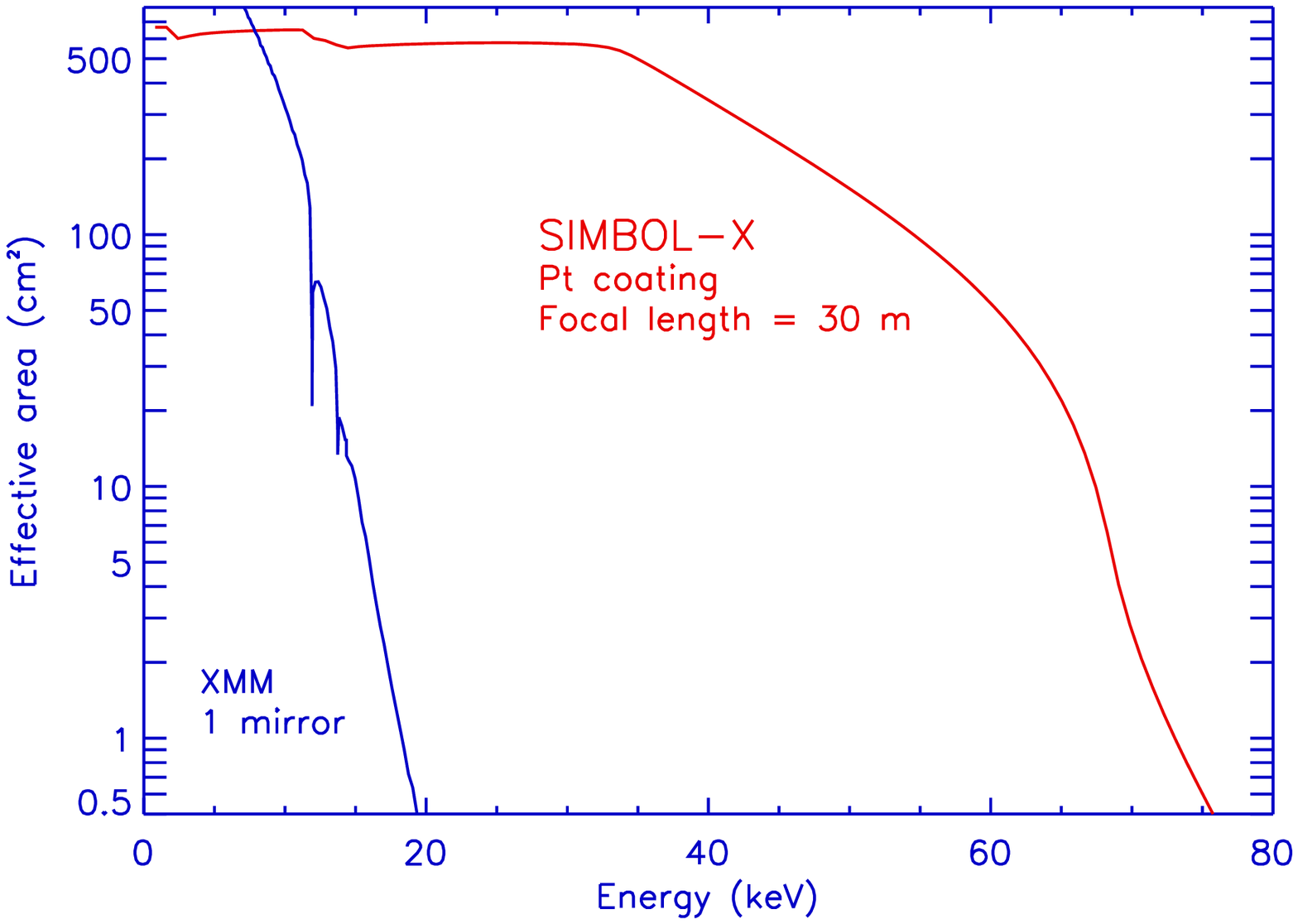}
\vspace{-1.0cm}
\caption{Mirror effective area in the baseline configuration, compared 
to that of one XMM--Newton mirror.}
\label{fig:mirror}
\end{minipage}
\end{figure}

SIMBOL--X is built on the classical design of a Wolter~I optics 
focusing X--rays onto a focal plane detector system. 
The gain in maximum energy is achieved by having a
focal length of 30 metres, \ie 4 times that of XMM--Newton 
mirrors. Since this cannot fit in a single spacecraft, the mirror 
and the detectors will be flown on two separate spacecrafts, in a 
formation flying configuration, as sketched in Fig.~\ref{fig:fly}. We 
shortly detail below each part of this system.

The focusing optics will be a nested shells Wolter~I configuration mirror. 
Building on the experience acquired on Beppo--SAX, Jet--X, SWIFT, 
ABRIXAS and XMM--Newton mirrors, it will be made following the
Nickel electroforming replication method (\eg Citterio \etal 2001).
The current design is to have a 108 shells mirror, with an outer 
diameter of 70~cm (like one XMM--Newton module), and an angular 
resolution of 20 arcseconds of Half-Energy-Width. The coating will be 
Pt, in order to increase the high energy response w.r.t.\ a more 
classical Au coating. The focal length will be 30 metres.

Figure \ref{fig:mirror} shows the effective area as a function of 
energy, for the baseline design. It has roughly a constant value of 
600~cm$^2$ up to about 35~keV, before starting to decrease and fall 
below one cm$^2$ above more than 70~keV. The field of view will be 
of 6 arcminutes at 50~\% vignetting.

The focal plane detector system will be a 6~cm diameter spectro-imager
with 500~$\mu$m of maximum pixel size (to provide an oversampling of 
the mirror PSF) with a ``reasonable" spectral resolution below 10~keV 
for measuring lines, particularly that of Iron, and a 100~\% efficiency 
at high energy. Since this cannot be given by a single detector, the 
focal plane will combine a low energy detector, stopping photons up to 
$\sim$~15~keV, directly on top of high energy detector. This is 
completed by an optical blocking filter, and an active anticoincidence 
shield. The best option for the low energy detector today is a DEPFET 
based APS (\eg Holl 2002) which operates at room temperature, but thick 
depletion CCDs would also fit SIMBOL--X requirements. These detectors 
are currently under tests in the MPE Garching and Leicester 
University. The high energy detector will consist in 2~mm thick CdZnTe 
pixellated arrays (\eg Limousin 2002) currently under tests in CEA/SAp.

Both mirror and detector spacecrafts will be of the ``mini-satellite" 
class (500~kg max). To keep a constant image quality requires the 
following constraints on the spacecraft relative positionning : i) their
distance must be kept constant within 1~cm, ii) their positionning 
perpendicular to the optical axis must be kept within 1~cm, and monitored 
with a 0.5~mm accuracy, and iii) the angular stability must be better than 
1~arcmin, and monitored to 3~arcsec. In order to minimize the 
differential forces between the 2 spacecrafts, as well as to allow 
uninterrupted observations of variable sources, SIMBOL--X will 
be put in orbit around L2.

\section{Sensitivity}
\begin{figure}[ht]
\includegraphics[width=\linewidth]{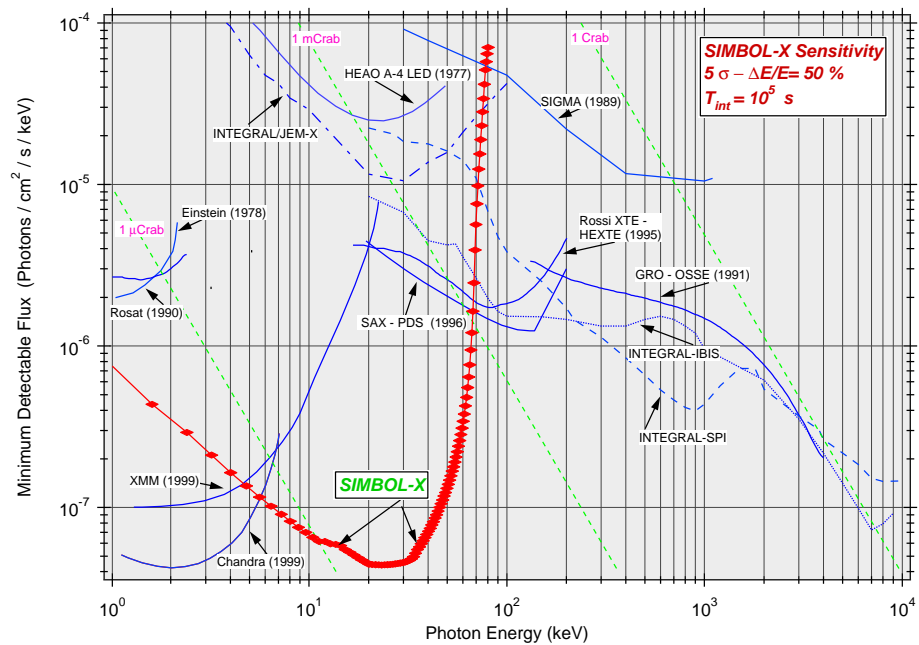}
\caption{SIMBOL--X sensitivity to point sources, compared to past and
present X and $\gamma$--ray telescopes.}
\label{fig:sensi}
\end{figure}

Figure~\ref{fig:sensi} presents the SIMBOL--X sensitivity to point 
sources, or equivalently to diffuse emission on a 1 arcmin 
diameter scale, compared to other instruments. SIMBOL--X curve was 
derived using the background spectrum of Ramsey (2001) for the HXT CdZnTe 
detector envisionned for Constellation~X (also at L2 position), and a 
model of the astrophysical diffuse background. As expected for an X--ray 
focusing telescope, the SIMBOL--X sensitivity curve has roughly the shape of the 
XMM--Newton and Chandra curves (which have no diffuse background 
component here) but is displaced by about a decade in energy. 
SIMBOL--X is $\sim$~100 times better than existing instruments 
in the 10 to 35~keV range, and has a sensitivity equivalent to 
INTEGRAL/IBIS at $\sim$ 70 keV.

We have also used this background model to simulate a number 
of observations that cannot be detailed here. We simply mention two 
examples to illustrate the SIMBOL--X sensitivity. On the supernovae 
remnant side, a detailed map of Cas~A above 20~keV can be done in 
100~ksec, and the spectrum of its brightest 1~arcmin$^2$ part is 
significant up to 50~keV. On the AGN jet side, the spectrum of the 
Pictor~A hot spot at 4~arcmin from the nucleus (well isolated with 
SIMBOL--X optics) can be significantly measured up to 40~keV in 
50~ks of observation.

\vspace{-0.6cm}
\section{Collaboration and schedule}
SIMBOL--X is a collaboration between France (CEA/Saclay, 
PI Institute, and the Grenoble and Meudon observatories), Italy with
Brera Observatory, Germany with MPE Garching, and United Kingdom 
with Leicester University. 

As SIMBOL--X does not involve new difficult technological development 
neither on the detector side nor on the mirror side, and as
the formation flying constraints are rather well within current 
investigations in this domain, a relatively short development time can 
be envisionned. Launching SIMBOL--X before the end of the decade 
would provide an excellent scientific preparation to the much larger
observatories Constellation~X and XEUS that are scheduled later. 

SIMBOL--X has been presented to the Astrophysics group of CNES, the 
french space agency, with a proposed launch date of early 2009. It has 
been recommanded in June 2002 for a phase A study. 



\begin{thebibliography}{}
\vspace{-0.2cm}    
\bibitem{Citterio2001}
Citterio, O., Conconi, P., Ghigo M., \etal, 2001, Proc.\ SPIE {\bf 4012}, 530.
\bibitem{Holl2002}
Holl, P., Fischer, P., Hartmann, R., \etal, 2002, Proc. SPIE 4851, in press, 2002
\bibitem{Limousin2002}
Limousin, O., 2002, Proc.\ of the 3rd Beaune conf., NIM A, in press
\bibitem{Ramsey2001}
Ramsey B.D., 2001, ``12th Int.\ Workshop on Room Temp.\ 
Semicon.\ X- and $\gamma$-Ray Det."
\end{thebibliography}
\end{document}